\documentclass[preprintnumbers,amsmath,amssymb,twocolumn,superscriptaddress]{revtex4}

\usepackage{amsmath}
\usepackage{amsfonts}
\usepackage{amssymb}
\usepackage{graphicx}
\usepackage{epsfig} 
\usepackage{color}

\newcommand{\mbf}[1]{\mathbf{#1}}
\newcommand{\mrm}[1]{\mathrm{#1}}
\newcommand{\mcl}[1]{\mathcal{#1}}

\begin{document}

\title{A one-dimensional dipole lattice model for water in narrow nanopores}

\author{J\"urgen K\"ofinger}
\affiliation{Faculty of Physics, University of Vienna,
Boltzmanngasse 5 , 1090 Vienna, Austria }
\author{Gerhard Hummer}
\affiliation{Laboratory of Chemical Physics, Building 5, National Institute of Diabetes and Digestive and Kidney Diseases, National Institutes of Health, Bethesda, Maryland, 20892-0520}
\author{Christoph Dellago}
\affiliation{Faculty of Physics, University of Vienna, Boltzmanngasse 5 , 1090 Vienna, Austria }

\date{\today}

\begin{abstract}

We present a recently developed one-dimensional dipole lattice model that accurately captures the key properties of water in narrow nanopores. For this model, we derive three equivalent representations of the Hamiltonian that together yield a transparent physical picture of the energetics of the water chain and permit efficient computer simulations. In the charge representation, the Hamiltonian consists of nearest-neighbor interactions and Coulomb-like interactions of effective charges at the ends of dipole ordered segments. Approximations based on the charge picture shed light on the influence of the Coulomb-like interactions on the structure of nanopore water. We use Monte Carlo simulations to study the system behavior of the full Hamiltonian and its approximations as a function of chemical potential and system size and investigate the bimodal character of the density distribution occurring at small system sizes.
\end{abstract}

\maketitle

\section{Introduction}
Confinement of water into molecularly narrow pores dramatically influences its structure and dynamics. For instance, the water flow through single wall carbon nanotubes is orders of magnitude larger than that expected from continuum hydrodynamics \cite{HoltBakajinScience2006}, with water fluxes comparable with those of transmembrane proteins \cite{FornasieroBakajinPNAS2008}. In carbon nanotubes with sub-nanometer diameters, water forms wires that provide  ideal routes for proton conduction \cite{DellagoHummerPRL2003,DellagoHummerPRL2006,HummerMolPhys2007}. Due to these properties, water-filled carbon nanotubes are promising building blocks  for future high flux membranes \cite{KalraHummerPNAS2003,HoltBakajinNanoLett2004,KreuerSchusterChemRev2004,HoltBakajinScience2006,LiuCrudenJPCB2006}, desalination systems \cite{KalraHummerPNAS2003,PeterHummerBiophysJ2005,FornasieroBakajinPNAS2008,CorryJPCB2009}, and nanofluidic devices \cite{HoltBakajinNanoLett2004}, as well as models for biological water and ion transport  \cite{Hille2001,ZhuSchultenBiophysJ2003}. 

Several recent computer simulation studies have focused on the structure and energetics as well as the transport properties of water in the interior of narrow carbon nanotubes \cite{HummerNoworytaNature2001,DellagoHummerPRL2003,DellagoHummerPRL2006,HummerMolPhys2007}. These simulations indicate that in a water bath at ambient conditions such nanotube channels  fill with water despite the hydrophobic nature of their walls \cite{HummerNoworytaNature2001}, a prediction which has been confirmed by experiment \cite{KolesnikovBurnhamPRL2004,NaguibYoshimuraNanoLett2004}. For sufficiently small nanotube diameters, the smooth inner tube wall forces the water molecules to stay near the tube axis such that they form a hydrogen bonded and orientationally ordered single file chain. Remarkably, these chains remain ordered to almost macroscopic length scales of  $\gtrsim 0.1$ mm due to the relatively high energy of orientational defects \cite{KoefingerDellagoPNAS2008}. For larger system sizes these defects destroy the order and no true order/disorder phase transitions occurs in the thermodynamic limit as required by theory \cite{RuelleCommMathPhys1968,LuijtenBloetePRB1997}. 

Due to the reduced mobility of the water molecules in the one-dimensional confinement, single file water chains in narrow pores can be modelled using lattice models with discrete degrees of freedom. Such simplified models capture the essential physics of diverse phenomena ranging from tube filling to the protonic conduction and water diffusion \cite{DellagoHummerPRL2003,MaibaumChandlerJPCB2003,ChouJPAMathGen2002,ChouBiophysJ2004}. Recently, we have introduced such a one-dimensional lattice model, in which water molecules are represented as as point dipoles oriented either parallel or orthogonal to the tube axis \cite{DellagoHummerPRL2003,KoefingerDellagoPNAS2008}. In contrast to other lattice models, our dipole model quantitatively reproduces the structure of quasi one-dimensional water in the tube interior including the formation of  defects and their interactions. Moreover, simulations of this model are computationally inexpensive such that studies of large systems are possible that would not be feasible otherwise. 

Here, we present a detailed derivation of our lattice model and its various different, mathematically equivalent, representations. In this model, dipoles are arranged on a regular 1d-lattice and interact via $1/r^3$ dipole-dipole interactions. This {\it dipole picture} can be simplified by grouping domains with equal orientation into segments. In the resulting {\it segment picture}, the total energy of the system is written as a sum of the internal energies of the segments and their interactions, which are of the dipole-dipole type. As we shall see, this segment picture is especially useful for the formulation of Monte Carlo moves that satisfies the configurational constraints dictated by the model.
Resummation of the internal energy of the ordered segments finally leads to the  {\it charge picture}, in which the total energy is expressed as a sum of Coulomb-like interactions of effective charges placed at the endpoints of the ordered segments. 
In this physically appealing picture the Coulomb-like interactions account for all effective interactions of defects, chain ends, and protons.
Because of its reduced computational complexity, the charge representation permits simulations of tubes of macroscopic length and investigations of the approach of the thermodynamic limit. The charge picture lends itself to approximations in which the long-range interactions of the effective charges are neglected and which allow the investigation of the role of the Coulomb-like interactions. In these approximations the filling of the tube is reproduced correctly, but the defect number at the filling transition is not. For small system sizes the particle number distribution function is bimodal with peaks at low and high densities and is captured nicely by the approximations. However, we find that neglecting Coulomb-like interactions qualitatively alters the form of the low density peak.

The remainder of this paper is organized as follows. In Sec.\ \ref{MODEL} we develop the dipole model and derive the two other equivalent formulations of the Hamiltonian. One of these, the charge picture, is then used to obtain approximations of the Hamiltonian. In Sec.\ \ref{SIMMET} we introduce the simulation methods and in Sec.\ \ref{PARAM} we discuss the parameterization. The filling/emptying transition and the bistability of the particle number distribution are presented and discussed in the context of the approximations in Sec.\ \ref{RESULTS}. The paper concludes with a discussion in Sec. \ref{DISCUSS}.

\section{Model}

\label{MODEL}
\begin{figure}[htb]
\epsfig{file=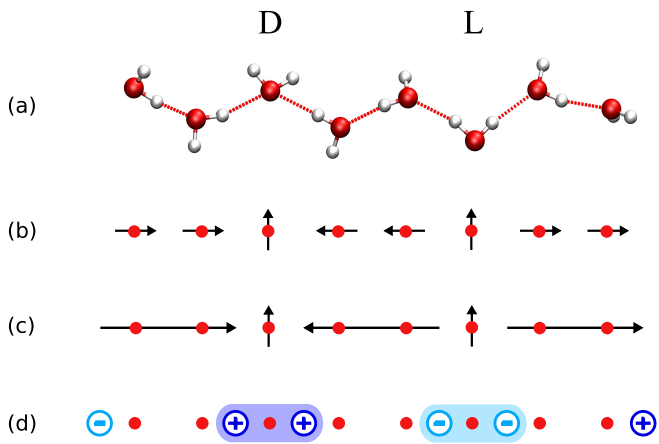, scale=0.5, clip}
\caption{ \label{PIC}
1D water wires in nanopores. (a) Chain configuration with a D-defect and an L-defect, and the corresponding lattice model in the (b) dipole representation, (c) the segment representation, and (d) the effective charge representation. In the lattice model, sites are marked by filled circles, dipoles are represented by short arrows, segments by long arrows, and effective charges by circles with their sign at the center. 
(a) The D-defect molecule accepts two hydrogen bonds from the two neighboring water molecules; in contrast, the L-defect molecule donates two hydrogen bonds.
(b) The next neighbor dipoles of the defect sites point to the defect site for the D-defect and away from it for the L-defect. 
(c) The configurational energy of the water wire due to the dipoles is determined by the length, orientation, and distance of segments. 
(d) In the effective charge representation, the segments are replaced by charges at the their ends with signs according to their orientations. As a consequence, defects are formed by pairs of equal charges which are positive for the D- and negative for the L defect.}

\end{figure}

Water in nanopores forms hydrogen-bonded chains of water molecules. Such a chain is dipole ordered if all molecules accept a hydrogen bond from the neighboring molecule on the left and donate a hydrogen bond to a molecule on the right (or the other way round). This order is destroyed by orientational defects. Note, however, that hydrogen-bond defects within a contiguous chain preserve the total number of hydrogen bonds.
Figure \ref{PIC}(a) shows a chain of water molecules that consists of three ordered segments connected by defect molecules. The segments consist of two molecules each and are orientationally ordered. The D-defect connects two segments pointing towards each other and an L-defect connects two segments pointing away from each other. In contrast to molecules within ordered segments which donate a single hydrogen bond and accept a single hydrogen bond, the D-defect molecule accepts two hydrogen bonds without donating any and the L-defect donates two hydrogen bonds without accepting any. This also means, that defects cannot be located at chain ends. Another configurational constraint is that defects can not be located next to each other, because the corresponding molecular configurations are unstable and lead to immediate recombination of the defects. Note, that the defect structures shown in Fig.\ \ref{PIC}(a) are the most typical configurations, but others can occur as well.

The free energetics of such a chain of water molecules are captured with great accuracy by a one-dimensional dipole lattice model. Molecular simulations show that due to the hydrogen bonds the water molecules are on average located on the sites of a one-dimensional lattice and that the long-range interaction of water molecules in an ordered chain is given by the dipole-dipole interaction. The magnitude of the dipoles is given by the average of the component of the dipole moment along the tube axis of a water molecule in an ordered chain. As a consequence, in the dipole model an ordered segment consist of equally oriented dipoles parallel to the tube axis, located on the site of a one-dimensional lattice [Fig.\ \ref{PIC}(b)]. The dipole moments of defect molecules are on average perpendicular to the tube axis and we only include their next-neighbor interactions.

On this basis, we can formulate an effective Hamiltonian with a reduced number of degrees of freedom.
This Hamiltonian describes the free energetics of arbitrarily filled tubes that in general consist of ordered or disordered hydrogen-bonded chains of water molecules (or fragments), with gaps between them.

Let us assume that our lattice has $N$ sites.
The lattice spacing is $a$ and the dipole moment $p$.
A dipole located on site $\nu$ has a direction $\sigma_\nu=1$ if it points ``up'', and $\sigma_\nu=-1$ if it points ``down'' the tube axis. 
The interaction potential of two dipoles located on sites $\nu$ and $\mu$ separated by a distance $d_{\nu\mu}=|\nu-\mu|a\neq0$ is given by 
\begin{equation}
\label{DIPDIPINT}
\phi_{\nu\mu} = -\varepsilon\frac{\sigma_\nu \sigma_\mu}{|\nu-\mu|^3} \quad .
\end{equation}
If a site $\nu$ carries a defect or if it is empty, we assign this site $\sigma_\nu=0$.
In these cases Eq.~(\ref{DIPDIPINT}) remains valid as  $\phi_{\nu\mu}=0$.
Here, we introduced the energy $ \varepsilon = \frac{2}{4\pi\varepsilon_0}\frac{p^2}{a ^3} $, where $\varepsilon_0$ is the dielectric constant, that sets the scale for the dipole-dipole interaction.
In the following, we use reduced units, i.e., $\varepsilon$ as unit for the energy, $a$ as unit for the length, and $p$ as unit for the dipole moment.

\subsection{Dipole picture}

Molecular simulations show that the interaction energy of next neighbor molecules, the so-called {\it contact energy}, is different from the dipole-dipole interaction energy of next neighbors, which is given by $\phi_{\nu,\nu+1}=-1$ for $\sigma_\nu \sigma_{\nu+1}=1$.
It will be useful to include the next neighbor dipole-dipole interaction in the Hamiltonian and correct for it to get the right contact energies. Let the configuration consist of $n$ water molecules (occupied sites), $n_\mrm{c}$ hydrogen bonded chains of molecules (fragments), and $n_\mrm{d}$ defects. We add the contact energy $E_\mrm{c}$ for each of the $(n-n_\mrm{c})$ hydrogen bonds and subtract the next neighbor dipole interaction for the $(n-n_\mrm{c}-2 n_\mrm{d})$ next neighbor pairs of parallel dipoles. 
This leads to the following effective Hamiltonian for water in nanopores
\begin{equation}
        \label{HD}
        H = \sum_{\nu=1}^{N-1} \sum_{\mu=\nu+1}^{N} \phi_{\nu\mu}
          +(n-n_\mrm{c})(1+E_\mrm{c})-2 n_\mrm{d}+n_\mrm{c} S_\mrm{c}\, , 
\end{equation}
where the double sum extends over all pairs of sites.
$S_\mrm{c}$ is an entropic contribution that accounts for the different contributions to the phase space volume of molecules at the chain ends and at defect sites compared to molecules within the chain. These different contributions are related to the number of dangling OH bonds (see Sec.\ \ref{PARAM}).
As we will see, this Hamiltonian is just one of three equivalent descriptions of the system. 
We will refer to this way of calculating the Hamiltonian as the {\it dipole picture} [see Fig.\ \ref{PIC}(b)], as we sum over all dipole-dipole interactions.

\subsection{Segment picture}

We can also formulate the Hamiltonian of Eq.\ (\ref{HD}) in terms of the $n_\mrm{s}= n_\mrm{c}+n_\mrm{d}$ segments, numbered from left to right, leading to the so-called {\it segment picture} [see Fig.\ \ref{PIC}(c)]. 

For this purpose, we need the {\it dipolar internal energy} of a segment, which stems from the interaction of the dipoles within a segment, and the {\it dipolar interaction energy} of two segments, which stems from the interaction of dipoles belonging to two different segments.
For brevity we will drop the term ``dipolar'' and only speak of internal and interaction energies of segments in the following.

The beginning of the segment with index $i$ is given by its coordinate $x_i$ and its length is denoted by $l_i$. 
The coordinates of the dipoles of segment $i$ are given by $x_i+n-1/2$ with $1\leq n \leq l_i$.
As all dipoles within a segment $i$ have the same direction, i.e., $\sigma_\nu=s_i$ for all dipoles within the segment $i$, we assign each segment a direction $s_i=\pm1$. 
The internal energy $E(l_i)$ of a segment $i$ is given by the sum over all pair interactions of dipoles $\nu$ within the segment, i.e.\ ,
\begin{equation}
\label{INTERNAL1}
E(l_i)=-\sum_{\nu=1}^{l_i-1}\sum_{\mu=\nu+1}^{l_i}\frac{1}{ |\mu-\nu|^{3}} \quad . 
\end{equation}
The interaction energy $I(l_i, l_j, s_i s_j, \Delta l_{ij})\equiv I_{ij}$ of segments $i$ and $j$, with $i<j$, is defined as   
\begin{equation}
        I_{ij} = - s_i s_j\sum_{\nu=1}^{l_i} \sum_{\mu=1}^{l_j} \frac{1}{(\mu-\nu+\Delta l_{ij}+l_i)^{3}} \quad ,
\end{equation}
with the size of the gap between the two segments given by $\Delta l_{ij}=x_j-(x_i+l_i)$.

In the segment picture we obtain for the Hamiltonian of a single chain  
\begin{eqnarray}
        \label{ETOT1}
         H &=& \sum_{i=1}^{n_\mrm{s}} E(l_i) + \sum_{i=1}^{n_\mrm{s}-1} \sum_{j=i+1}^{n_\mrm{s}} I_{ij} + \nonumber \\ 
        &+& (n-n_\mrm{c})(1+E_\mrm{c})-2 n_\mrm{d} +n_\mrm{c} S_\mrm{c}  \quad. 
\end{eqnarray}

The calculation of the energy according to Eq.~(\ref{ETOT1}) in the segment picture can be simplified considerably by deriving an explicit functional form for the internal energy and expressing the interaction energies in terms of the internal energy. 
Firstly, we can avoid the double sum in the calculation of the internal energy of Eq.\ (\ref{INTERNAL1}) by counting all pairs of dipoles separated by a certain distance. In a segment of length $l$, the interaction potential of dipoles separated by a distance $j$ with  $1\leq j < l$ appears $(l-j)$ times. This leads to
\begin{eqnarray}
E(l)&=& -\sum_{j=1}^{l-1} (l-j)j^{-3} = \\
        &=&\label{EL1}\sum_{j=1}^{l-1} j^{-2} - l \sum_{j=1}^{l-1} j^{-3} 
\end{eqnarray}
for Eq.\ (\ref{INTERNAL1}).

For $l\rightarrow\infty$, the two sums in Eq.\ (\ref{EL1}) the two sums can be expressed in terms of Riemann's zeta function \cite{AbramowitzStegun1965}  
\begin{equation}
\zeta(m)=\sum_{j=1}^\infty j^{-m} \quad .
\end{equation}
Accordingly, for long segments we can approximate the internal energy $E(l)$ as
\begin{equation}
\label{EAPPROX}
E_{l\gg 1}(l) = \zeta(2)-l \zeta(3) \quad .
\end{equation}
The difference,  $\Phi (l) =E_{l\gg 1}(l)-E(l) $ , between the above approximation and the exact internal energy is given by  
\begin{eqnarray}
  \Phi (l) &=&  \sum_{j=l}^\infty \frac{1}{j^2} - l \sum_{j=l}^\infty \frac{1}{j^3} 
\end{eqnarray}
and can be rewritten as 
\begin{eqnarray}
  \Phi (l) &=& \Psi'(l)+\frac{l}{2}\Psi''(l) \label{EDIFF} 
\end{eqnarray}
using the polygamma function \cite{AbramowitzStegun1965} 
\begin{equation}
\Psi^{(m)}(l)=(-1)^{m+1} m! \sum_{j=0}^\infty \frac{1}{(l+j)^{m+1}}\, .
\end{equation}
Hence, the internal energy of a segment of length $l$ can be written as the sum of a linear part, Eq.\ (\ref{EAPPROX}), and a non-linear part, Eq.\ (\ref{EDIFF}), leading to the exact expression   
\begin{equation}
\label{INTERNAL}
E(l)=\zeta(2)-l \zeta(3)-\Phi(l) \quad .
\end{equation}

\begin{figure}[t]
\epsfig{file=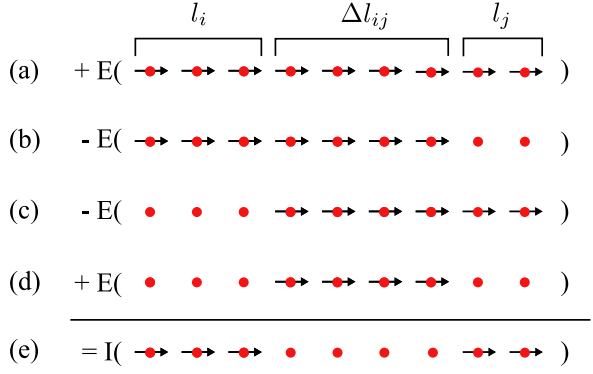, scale=0.4, clip}
\caption{ \label{FIG_INTER_ENERGY} Calculation of the interaction energy. The interaction energy $I$ of configuration (e), consisting of two chains of length $l_i$ and $l_j$, separated by a distance, $\Delta l_{ij}$, of four sites, can be calculated by subtracting the internal energies $E$ of configurations (b) and (c) from the internal energy of (a), and adding the internal energy of the configuration (d).}
\end{figure}

The next step towards a simpler energy calculation in the segment picture is to express the interaction energy of two segments $i$ and $j$ in terms of internal energies (see Fig.\ \ref{FIG_INTER_ENERGY}). 
As the directions of the segments only determine the sign of their interaction energy, we assume, for the moment, that the two segments of length $l_i$ and $l_j$ have the same direction, Fig.\ \ref{FIG_INTER_ENERGY} (e).   

In the following we use the simple fact that the internal energy $E(l)$ of a chain of length $l$ can be calculated from its parts of length $l'$ and $l''$, with $l=l'+l''$,  as
\begin{equation}
 E(l)=E(l')+E(l'')+I(l', l'',1,0)\quad .
\end{equation}
For brevity we drop $s_{ij}$ and $\Delta l_{ij}$ as arguments of $I(l_i, l_j, s_{ij}, \Delta_{ij})\equiv I(l_i, l_j)$ here.
Using the equation above the internal energy of a segment with length $l=l_i+l_j+\Delta l_{ij}$ [Fig.\ \ref{FIG_INTER_ENERGY} (a)] can be written as
\begin{eqnarray}
\label{EL}
E(l) &= &E(l_i)+E(l_j)+E(\Delta l_{ij})+\nonumber \\
&+& I(l_i,l_j)+I(l_i,\Delta l_{ij})+I(\Delta l_{ij},l_j)
\end{eqnarray}
To get the desired interaction energy, $I(l_i,l_j)$ on the right hand side of Eq.~(\ref{EL}),  we subtract the internal energies of a segment of length $l_i+\Delta l_{ij}$ and of a segment of length $\Delta l_{ij}+l_j$ [Fig.\ \ref{FIG_INTER_ENERGY} (b) and (c)] .
These two energies, given by 
\begin{eqnarray}
\label{EPARTS1}
 E(l_i+\Delta l_{ij})&=&E(l_i)+E(\Delta l_{ij})+I(l_i,\Delta l_{ij}) \\ 
\label{EPARTS2}
 E(\Delta l_{ij}+l_j)&=&E(\Delta l_{ij})+E(l_j)+I(\Delta l_{ij}, l_j) 
\end{eqnarray}
include the internal energies of the two segments $i$  and $j$, and the interaction energies of the segment $i$ and $j$ with the segment of length $\Delta l_{ij}$ separating them.
They also include twice the internal energy $E(\Delta l_{ij})$ of the segment in the middle. To correct for this double subtraction, we have to add this energy once, Fig.\ \ref{FIG_INTER_ENERGY} (d).
As the linear terms of Eq.\ (\ref{INTERNAL}) cancel we obtain for the interaction energy  
\begin{eqnarray}
\label{INTERACTION}
I_{ij} &=& -s_i s_j \left[\Phi(l_i+l_j+\Delta l_{ij}) + \Phi(\Delta l_{ij}) + \right.\nonumber \\
&& \left .- \Phi(l_i+\Delta l_{ij})- \Phi(l_j+\Delta l_{ij})\right] \quad .
\end{eqnarray}
Note that this expression for the interaction energy is also valid for continuous gap distances $\Delta l_{ij}\geq 0$.

This result can also be obtained by simply inserting the expressions for the interaction energies $I(l_i,\Delta l_{ij})$ and $I(\Delta l_{ij}, l_j)$ following from  Eqs.~(\ref{EPARTS1}) and (\ref{EPARTS2}),
\begin{eqnarray}
\label{IPARTS1}
 I(l_i,\Delta l_{ij})&=&E(l_i+\Delta l_{ij})-E(l_i)-E(\Delta l_{ij}) \\ 
\label{IPARTS2}
 I(\Delta l_{ij}, l_j)&=&E(\Delta l_{ij}+l_j)-E(\Delta l_{ij})-E(l_j) 
\end{eqnarray}
to Eq.~(\ref{EL}).

Using Eqs.\ (\ref{INTERNAL}) and (\ref{INTERACTION}) we can write the total energy in the segment picture, Eq.\ (\ref{ETOT1}), as
\begin{eqnarray}
        \label{HCHAIN2}
        H &=&-\sum_{i=1}^{n_\mrm{s}-1}\sum_{j=i+1}^{n_\mrm{s}} s_i s_j \left[\Phi(l_i+l_j+\Delta l_{ij}) + \Phi(\Delta l_{ij}) + \right.\nonumber \\
&& \left .- \Phi(l_i+\Delta l_{ij})- \Phi(l_j+\Delta l_{ij})\right] - \sum_{i=1}^{n_\mrm{s}} \Phi(l_i) +\nonumber \\
        &&   + n_\mrm{d} c_\mrm{d}+n_\mrm{c} c_\mrm{c}+n c\,  .         
\end{eqnarray}
This Hamiltonian is linear in the number of defects $n_\mrm{d}$, the number of fragments $n_\mrm{c}$, and the number of occupied sites $n$. The corresponding coefficients $c_\mrm{d}$, $c_\mrm{f}$, and $c$, which depend on the contact energy $E_\mrm{c}$ and the entropic contribution $S_\mrm{c}$, are given by
\begin{eqnarray}
 c_\mrm{d} &=& \zeta(2)+\zeta(3)-2\quad,  \\
 c_\mrm{c} &=& \zeta(2)-1-E_\mrm{c}+S_\mrm{c}\quad,  \\
 c &=& 1+E_\mrm{c}-\zeta(3)\quad .
\end{eqnarray}

\subsection{Charge picture}

We can use this expression for the Hamiltonian in the {\it segment picture} to derive another equivalent description of the system, i.e.\ the so-called {\it charge picture} [see Fig.\ \ref{PIC}(d)].
The first few terms of the series expansion of the non-linear part of the internal energy,
\begin{equation}
\Phi(l)\approx\frac{1}{2 l}-\frac{1}{12} \left(\frac{1}{l}\right)^3+\frac{1}{20} \left(\frac{1}{l}\right)^5-\mcl{O}\left[\left(\frac{1}{l}\right)^7\right]\, , 
\end{equation}
show that for large chain lengths $l$ the non-linear part is proportional to $1/l$, i.e., Coulombic.
Thus, we replace all segments by charges at their beginnings and their ends according to their orientations. As each segment carries two charges, the indices of the charges are given by $2i-1$ for the charge at the beginning and $2i$ for the charge at the ends of segment $i$. The charge at the beginning of segment $i$ is $q_{2i-1}=-s_i$ and the charge at the end is $q_{2i}=s_i$. 
The coordinates of these charges are $z_{2i-1}=x_i$ and  $z_{2i}=x_i+l_i$.
The interaction potential of two charges $q_m$ and $q_n$ separated by a distance $z_{mn}=|z_n-z_m|$ is given by 
$\Phi_\mrm{c}(z_{mn}) = q_m q_n \Phi(z_{mn})$. This interaction is Coulombic in character for large distances and we can rewrite this interaction as 
\begin{equation}
\varepsilon \Phi_\mrm{c}(z_{mn})\approx \varepsilon \frac{q_m q_n} {2 z_{mn}} = \frac{1}{4 \pi \varepsilon_0}\frac{Q_m Q_n} { z_{mn} a}. 
\end{equation}
with the magnitude of the charges then given by $|Q_m|=p/a$.
\begin{figure}[h]
        \epsfig{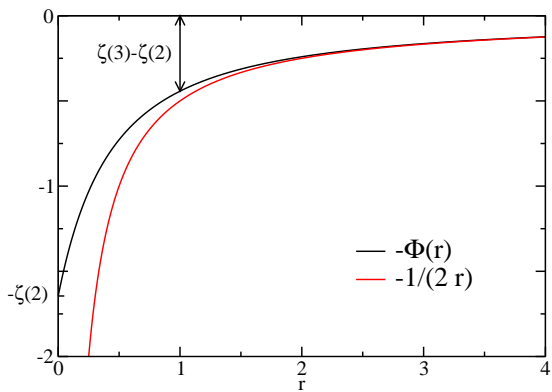}
\caption{\label{COULOMB}Comparison of the Coulomb interaction with the non-linear part of the internal energy. While the Coulomb interaction diverges for $r\to0$, $\Phi(0)=\zeta(2)$.}
\end{figure}
Figure \ref{COULOMB} shows a comparison of this Coulomb-like interaction, $\Phi(r)$, with the Coulomb interaction $1/r$.
The main difference is that for a distance $r\rightarrow 0$ the Coulomb-like interaction converges to a finite value, $\Phi(0)=\zeta(2)$, whereas the Coulomb-interaction diverges to $-
\infty$. Note that for all possible distances, $z_{mn}\geq 1$, the difference is small.

Rearranging the Hamiltonian in the segment picture given by Eq.~(\ref{HCHAIN2}) we obtain the Hamiltonian in the {\it charge picture} as
\begin{eqnarray}
        \label{ETOT_CHARGE}
        H &=&\sum_{m=1}^{2 n_\mrm{s}-1}\sum_{n=m+1}^{2 n_\mrm{s}} q_m q_n \Phi(z_{mn}) + \nonumber \\
        && + n_\mrm{d} c_\mrm{d}+n_\mrm{c} c_\mrm{c}+n c\,  .
\end{eqnarray}
The sum in Eq.\ (\ref{ETOT_CHARGE}) includes the interaction of the charges belonging to the same defect, given by $\Phi(1)=\zeta(2)-\zeta(3)$, which occurs $n_\mrm{d}$ times. Introducing the defect excitation energy, i.e., the energy (the free energy in the molecular model) needed to introduce a single defect in an infinitely long chain, 
\begin{equation}
 E_\mcl{D}=c_\mrm{d}-\phi(1)=2\zeta(2)-2\, 
\end{equation}
we can write the Hamiltonian in the charge picture as
\begin{eqnarray}
        \label{ETOT_CHARGE_ED}
        H &=&{\sum_{m=1}^{2 n_\mrm{s}-1} \sum_{n=m+1}^{2 n_\mrm{s}}}' q_m q_n \Phi(z_{mn}) + \nonumber \\
        && + n_\mrm{d} E_\mcl{D}+n_\mrm{c} c_\mrm{c}+n c \quad .
\end{eqnarray}
where the prime indicates that the sum does not include the interaction of two charges belonging to the same defect with each other.
The coefficient $c_\mrm{c}$ corresponds to the energy needed to break an infinitely long chain, and move the resulting fragments infinitely far apart from each other. The coefficient $c$ is the energy required to add a single dipole to an infinitely long chain.
The Hamiltonian in the charge picture depends on the charge positions, $z_j$, the number of particles $n$, the number of chains $n_\mrm{c}$, and the number of defects $n_\mrm{d}$. We note that these quantities are not sufficient to specify a configuration of the dipole model unambiguously.

The Hamiltonian in the charge picture [Eq.~(\ref{ETOT_CHARGE_ED})] hightlights the Coulomb-like effective interactions of defects and chain ends. L- and D- defects attract each other Coulombically whereas defects of the same kind repel each other. Defects next to a chain end are always attracted by the charge at this end.

\subsection{Approximate representations}

If the distances between charges are large we should be able to neglect their Coulomb-like interactions that decay as $1/r$. 
Then, only the linear terms of Eq.\ (\ref{ETOT_CHARGE_ED}) remain. The Hamiltonian in this simplest approximation, which we refer to as the no-charge-approximation (NCA), is given by 
\begin{equation}
\mcl{H}_0  = n c +  n_\mrm{c} c_\mrm{c} +n_\mrm{d} E_\mcl{D}\, .
\end{equation}

For small  distances of effective charges the Coulomb-interaction is strong and cannot be neglected. 
This is the case for short chains and segments which carry charges at their ends. In particular,  
chains and segments of length one are represented by two charges of opposite sign which are separated by one lattice constant only.  
At the next higher level of approximation, we add the interaction energy $-\Phi(1)=-\zeta(2)+\zeta(3)$ between the charge pairs associated with each of the $n_\mrm{I}$ segments of length one.  The resulting Hamiltonian is
\begin{equation}
\mcl{H}_1  = n c +  n_\mrm{c} c_\mrm{c} +n_\mrm{d} E_\mcl{D} +n_\mrm{I}[-\zeta(2)+\zeta(3)]\quad .
\end{equation}
We will refer to this Hamiltonian as the singlet-charge-approximation (SCA) since we include the interaction of charge pairs associated with single dipoles.

One main feature of these approximations compared to the full Hamiltonian is that the approximations can be written as spin models (with three spin states for the NCA and four spin states for the SCA) with nearest-neighbor interactions only.

The NCA and SCA do not depend on the length, positions, or orientations of the segments. States specified by the particle number $n$ and $\mbf{n}_0=\{ n_\mrm{c}, n_\mrm{d}\}$ for the NCA, and by $n$ and  $\mbf{n}_1=\{ n_\mrm{c}, n_\mrm{d}, n_\mrm{I}\}$ for the SCA are thus degenerate.
Calculating the degeneracy of these states requires to count the number of states for these approximations as a function of these variables. For the NCA the number of states,  $\Gamma_0(N, n, \mbf{n}_0)$, is given by 
\begin{eqnarray}
\label{GAMMA0}
\Gamma_0 &=&  2^{n_\mrm{c}}{ n - n_\mrm{d}-1 \choose n_\mrm{s}-1}{n_\mrm{s}-1 \choose n_\mrm{c}-1}{N+1-n \choose n_\mrm{c}}
\end{eqnarray}
 with the number of segments, $n_\mrm{s}=n_\mrm{c}+n_\mrm{d}$, being a function of the defect number and the chain number. For the SCA the number of states,  $\Gamma_1(N,n, \mbf{n}_1)$, is given by 
\begin{eqnarray}
\label{GAMMA1}
\Gamma_1 &=&  2^{n_\mrm{c}}{ n - 2 n_\mrm{s}+ n_\mrm{c}-1 \choose n_\mrm{s}-n_\mrm{I}-1} {n_\mrm{s} \choose n_\mrm{I}}{n_\mrm{s}-1 \choose n_\mrm{c}-1} \nonumber \\
&\times& {N+1-n \choose n_\mrm{c}}
\end{eqnarray}
for $n>0$. If $n_\mrm{s}=n_\mrm{c}=n$ then $\Gamma_0=\Gamma_1=2^{n}{N+1-n \choose n}$.
For a derivation of these equations and see App.~\ref{app:NRSTATES}.

\subsection{Proton defects}

Up to now we only considered chains of intact water molecules. If an excess proton is introduced into the system, it forms a hydronium ion consisting of one oxygen atom and three hydrogen atoms. In a single-file chain, this hydronium ion donates two hydrogen bonds without accepting any. Thus,  structurally the hydronium ion corresponds to an L-defect with an additional proton \cite{DellagoHummerPRL2003}. In the charge picture, the effective interaction of this protonated L-defect is given by the sum of the effective interaction of an L-defect and the effective interaction of the proton with defects and chain ends. 

In Ref.~\cite{DellagoHummerPRL2003} an approximate expression for this effective interaction was derived.
For completeness we show next how the effective interaction of an additional proton located on an L-defect is included in the charge picture. 
In the dipole model the interaction of a proton located on an L-defect at site $i$ and a dipole with direction  $\sigma_j=\pm1$ at site $j$ is given by 
\begin{equation}
W_{ij} = \frac{1}{4 \pi \varepsilon_0}\frac{\sigma_j p e (j-i)}{ a^2 |j-i|^3}= \varepsilon'  \frac{\sigma_j(j-i)}{|j-i|^{3}}
\end{equation}
where $e$ is the elementary charge and $\varepsilon'=\varepsilon a e /(2 p)$ the unit of the energy for the rest of this subsection. 

Using the ideas of the derivation of the segment and the charge picture, we calculate the interaction energy of a proton and an ordered segment with $l$ dipoles of relative orientation $s=\sigma_j (j-i)/|j-i|$ with respect to the position of the proton, i.e., $s=1$ ($s=-1$) for dipoles pointing away (towards) the proton. Let us say the proton is located in the origin at the site with index zero, the first dipole of the segment is located at $j_1$, and the last at $j_2=j_1+l-1$. Using the polygamma function we obtain 
\begin{eqnarray}
 W(j_1,l) &=& \sum_{j=j_1}^{j_2} W_{ij}= s \sum_{j=j_1}^{j_2} j^{-2} = \\
&=& s \sum_{j=j_1}^{\infty} j^{-2} - s \sum_{j=j_2+1}^{\infty} j^{-2}= \\
&=& s \left[-\Psi'(j_1)+\Psi'(j_1+l)\right]  \, .
\end{eqnarray}
for the interaction energy of the proton with the segment.
Rewriting this interaction energy as a function of the distance $x=|j_1-i|+1/2$ of the beginning of the segment from the proton we get
\begin{eqnarray}
 W(x,l) &=& s\left[-\Psi'(x+1/2)+\Psi'(x+l+1/2)\right]  =\nonumber \\
&=& s \left[\bar{\Phi}\left(x\right)+\bar{\Phi}\left(x+l\right)\right]
\end{eqnarray}
with a Coulomb-like interaction given by 
\begin{equation}
 \bar{\Phi}(x)=\Psi'\left(x+\frac{1}{2}\right)\approx \frac{1}{x}-\frac{1}{12} \left(\frac{1}{x}\right)^3+\mathcal{O}\left({x}^{-5 \\}\right)\, .
\end{equation}
Thus, the interaction energy of a proton with the dipoles of a segment is equal to a Coulomb-like interaction of the proton with charges at the beginning and the end of the segment.  For a protonated L-defect in an infinitely long chain we obtain a proton energy of $E_\mrm{p}= -2\zeta(2)$, stemming from the two charges next to the defect.

For large distances, the interaction energy of a hydronium ion with a defect can be written as  
\begin{equation}
 \Phi_\mrm{PL}(z)=\frac{1}{4 \pi \varepsilon_0}\frac{\mp Q}{a z} \left(e-Q\right)
\end{equation}
where $Q=2 p/a$ is the magnitude of the total effective charge of a defect. The plus sign is valid for the interaction with a D-defect and the minus sign for that with an L-defect. Since the proton charge is larger than  $Q$, a hydronium ion is repelled by D-defects and attracted to L-defects, in agreement with Ref.~\cite{DellagoHummerPRL2003}. Also, the hydronium interacts repulsively with the endpoints of an otherwise ordered water chain. As a consequence, the preferred position of an excess proton in an isolated water wire is at the chain center.  

\section{Monte Carlo Simulation}
\label{SIMMET}

Next, we examine the thermodynamic behavior of water in nanopores in contact with a heat bath and a particle reservoir. Thus, we perform canonical and grand-canonical Monte Carlo simulations of the dipole model. 

For the Monte Carlo simulations of the dipole model we describe configurations in the segment picture in which gaps of unoccupied sites are represented as segments with $s=0$. Thus, a configuration consists of occupied sections (segments) and empty sections (gaps). (In the following we use the term ``segments'' only for segments of chains and the term ``section'' for segments of chains and empty gaps.) Between segments one finds either a defect or an empty section.
 The boundary conditions are given by two empty gaps at the beginning and the end of the lattice (see App.~\ref{app:NRSTATES}). A configuration is unambiguously defined by the beginning $x_i$, the length $l_i$, and the value of $s_i$ for each section $i$. Defects are located between next neighbor sections with $s\neq0$. Including the empty end sections, there are $n_\mrm{0}=n_\mrm{c}+1$ sections with $s=0$ corresponding to gaps. 

With a configuration given by $\mathcal{C}_i=\{\{x_j,l_j,s_j\}: 1\leq j\leq n_\mrm{s}+n_\mrm{0}\}$
the canonical partition function can be written as 
\begin{equation}
 Z_N (\beta, n) = \sum_{\{\mathcal{C}_i\}} e^{-\beta H(\mathcal{C}_i)}\, .
\end{equation}
The grand canonical partition function is given by 
\begin{equation}
 \Xi_N (\beta, z) = 1 + \sum_{n=1}^{\infty} Z_N (\beta, n) z^n \, .
\end{equation}
The fugacity is defined as $z=e^{\beta \mu}$, where $\mu$ is the chemical potential and $
\beta=1/(k_\mrm{B} T)$ the reciprocal temperature with $k_\mrm{B}$ being Boltzmann's constant. 

To enhance the sampling, we use non-local Monte Carlo moves.  Monte Carlo simulations with only local moves, in which individual dipoles are flipped, are inefficient for large systems. Here, we apply efficient non-local trial moves with asymmetric generation probabilities for which we have to correct in the acceptance probability.
These trial moves change the lengths of sections and their orientations for the generation and recombination of defects, for the displacement of defects and chains, and for the insertion and deletion of particles. For details see App.~\ref{TRIAL}.

We also perform Monte Carlo simulations for the NCA and SCA for which the canonical partition functions are given by
\begin{equation}
\label{CAN_Z_SRA}
 \mcl{Z}_\mrm{N}^{(i)}(n)= \sum_{\{\mathcal{C}_j\}}e^{-\beta \mcl{H}_i(n,\mbf{n}_i)}
\end{equation}
with $i=0$ for the NCA and $i=1$ for the SCA, and $\mbf{n}_i$ implicitly depending on the configurations $\mathcal{C}_j$.
Using the degeneracies given by Eqs.~(\ref{GAMMA0}) and (\ref{GAMMA1}), Eq.~(\ref{CAN_Z_SRA}) can be rewritten as
\begin{equation}
\label{CAN_Z_SRA_too}
 \mcl{Z}_\mrm{N}^{(i)}(n)= \sum_{\mbf{n}_i} \Gamma_i(N, n, \mbf{n}_i) e^{-\beta \mcl{H}_i(n, \mbf{n}_i)} \, ,
\end{equation}
which permits us to formulate new effective ``Hamiltonians''
\begin{equation}
\label{HEFF}
 \mcl{H}_i'=\mcl{H}_i-T \ln \Gamma_i(N, n, \mbf{n}_i)
\end{equation}
with canonical partition functions given by
\begin{equation}
  \mcl{Z}^{(i)}_\mrm{N}(n)=\sum_{\mbf{n}_i}e^{-\beta \mcl{H}_i'(n, \mbf{n}_i)} \quad.
\end{equation}
We can calculate these partition functions either numerically by direct summation or perform Monte Carlo simulations in the space of $\mbf{n}_0=\{ n_\mrm{c}, n_\mrm{d}\}$ for the NCA and of  $\mbf{n}_1=\{ n_\mrm{c}, n_\mrm{d}, n_\mrm{I}\}$ for the SCA.  The applied trial moves simply increase or decrease the values of the variables of the Hamiltonian within the limits of the phase space given in App.~\ref{app:NRSTATES}.

To study the system behavior over a broad range of the chemical potential we use the Wang-Landau algorithm \cite{WangLandauPRE2001} with the particle number as order parameter to find a bias function $w(n)$ corresponding to the negative free energy as a function of the particle number. We use this function for a biased simulation \cite{FrenkelSmit2002} at the fugacity $z$, resulting in a flat histogram of the particle number, with the Hamiltonian in the biased system given by 
\begin{equation}
 H'=H-T w(n) \ln z 
\end{equation}
The output are samples of the total energy, chain number, defect number, number of particles, and the total dipole moment,  $\{E^{(i)}, n_\mrm{c}^{(i)}, n_\mrm{d}^{(i)}, n^{(i)}, D^{(i)}\}$. By unfolding the bias function $w(n)$ and reweighting \cite{FrenkelSmit2002} we obtain estimates for observables that are functions of the above quantities (see App.~\ref{BIAS}).

 \section{Parameterization}
\label{PARAM}

The dipole model was parameterized with and validated against detailed molecular simulations \cite{KoefingerDellagoPNAS2008}. 
The molecular model consists of a (6,6)-type carbon nanotube, filled with up to 100 water molecules, following Ref.~\cite{RasaiahHummerAnnRevPhysChem2008}. 
We used the TIP3P potential \cite{JorgensenKleinJCP1983} for the water-water interactions and a cylindro-symmetric potential for the tube-water interaction as in Refs.~\cite{RasaiahHummerAnnRevPhysChem2008,HummerNoworytaNature2001}.
Boundary effects are minimized by using tubes that are longer than the volume accessible to the water molecules.

The lattice spacing $a$, the dipole moment $p$, and the contact energy $E_\mrm{c}$ can be determined in a canonical Monte Carlo simulation of a single hydrogen bonded chain of water molecules in a nanopore. 

The lattice spacing is given by the average distance of neighboring molecules within a chain for which we obtain $a=2.65\ \mrm{\AA}$. The dipole moment is given by the average dipole moment of a water molecule in an ordered chain along the tube axis, $p = 1.9975\,\mrm{D}$. 
This determines the energy scale for the dipole-dipole interaction as $\varepsilon=2 p ^2 / (4\pi \varepsilon_0 a^3)= 25.8236$ kJ/mol, such that $\beta \varepsilon \approx 10.42$ at $T=298$ K. The contact energy $E_\mathrm{c} = -20.8$ kJ/mol is determined as the average interaction energy of two neighboring water molecules within a chain.

\begin{figure}[h]
        \epsfig{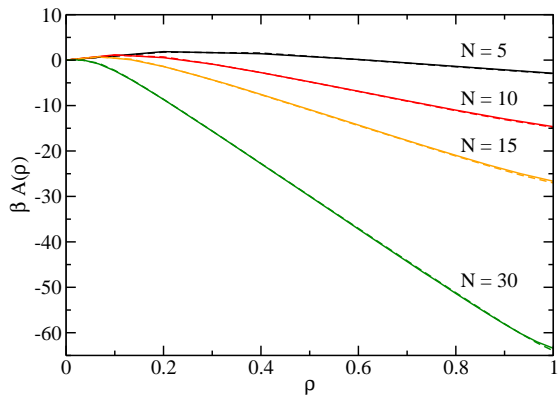}
 \caption{\label{TFE} The transfer free energy as a function of the density for different system sizes for the molecular model (solid lines) and the dipole model (dashed lines). The excellent agreement renders the curves for the different models nearly indistinguishable.  Results are for fugacity $z_0$.}
\end{figure}

To perform grand canonical simulations of the dipole model we have to determine the effective tube-water interaction. Since it couples to the particle number we can absorb it in the chemical potential. We also need the entropic contribution which couples to the number of chains. We determine both quantities by comparing the transfer free energies of the molecular model and the dipole model and tuning the fugacity and the entropic contribution to obtain agreement between these two models.
The transfer free energy, given by $\beta A(n) = -\ln [P(n)/P(0)]$, where $P(n)$ is the particle-occupancy distribution function, is the free energy needed to introduce $n$ water molecules into a tube of length $L$ at the reciprocal temperature $\beta$. 
We obtain for a tube of length $L=30 a$ in contact with a heat bath and a particle reservoir at ambient conditions, i.e., at room temperature and atmospheric pressure, a fugacity $z_0=0.000327$ and an entropic contribution $\beta S_\mrm{c}=-3.96$.

Figure~\ref{TFE} shows a comparison of the transfer free energies of the molecular and the dipole model for system sizes $N=5,\, 10,\, 15$, and $N=30$ as a function of the density $\rho=n/N$. 
Since the dipole model is a coarse-grained description of the molecular model, all states in the molecular model with an occupation number equal to or larger than the number of sites contribute to the completely filled state, i.e., to $P(N)$. 
The agreement is excellent as the curves for the molecular and the dipole model lie practically on top of each other. We observe that for small sizes the system shows two minima, one for the empty and one for the filled state. Thus, the particle number distribution function is bimodal, i.e., has two peaks  (see Sec.~\ref{BI}). For growing system size the minimum corresponding to the filled state gets deeper and the minimum corresponding to the empty state vanishes. 
\begin{figure}[h]
        \epsfig{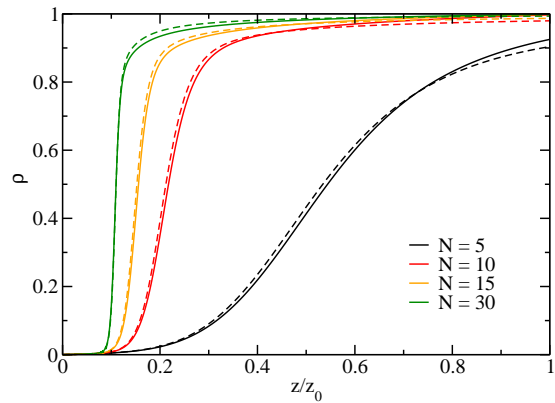}
\caption{\label{DENS} The particle density as a function of the relative fugacity.  The solid lines are results from molecular simulations and the dashed lines are results for the dipole model.}
\end{figure}

\begin{figure}[h]
        \epsfig{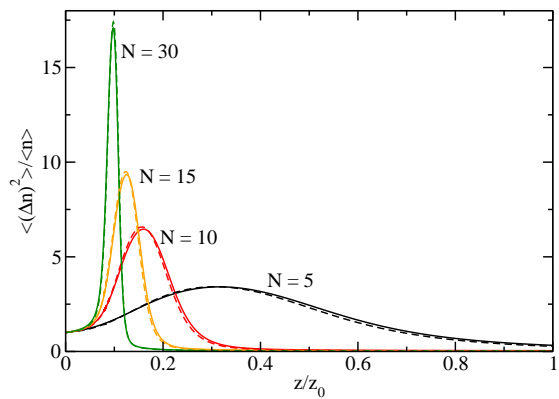}
\caption{\label{VAR} Particle number fluctuations as a function of the relative fugacity for different system sizes. The solid lines are results from molecular simulations and the dashed lines are results for the dipole model.}
\end{figure}

By reweighting the particle number distribution functions corresponding to the transfer free energies shown in Fig.~\ref{TFE} we obtain the adsorption isotherms and the relative particle number fluctuations as a function of the relative fugacity (Figs.~\ref{DENS} and \ref{VAR}, respectively). The relative fugacity is the actual fugacity divided by the fugacity of a system in contact with a heat and particle reservoir at room temperature and atmospheric pressure. For low fugacities the relative fugacity is equal to the relative humidity. For larger system sizes the density shown in Fig.~\ref{DENS} gets steeper at the filling transition, which moves to smaller fugacities. 

The relative variance, i.e., the variance of the particle number divided by the average particle number, is a measure for the fluctuation of the particle number.
For macroscopic volumes, the relative variance is related to the isothermal compressibility $\kappa_\mrm{T}$ via  $(\langle n^2 \rangle - \langle n \rangle^2) / \langle n \rangle=\rho k_\mrm{B}T \kappa_\mrm{T}$, where angled brackets denote ensemble averages.
The relative variance (Fig.~\ref{VAR}) has its maximum at the filling transition. The fluctuations become larger with increasing system size. The properties of the filling transition for increasing system size are discussed in detail in the Sec.~\ref{RESULTS}.

The excellent agreement of the dipole model and the molecular model for different system sizes supports the validity of the entropic contribution. It accounts for the different contributions to the phase space volume of water molecules according to the number of their dangling OH bonds. In an ordered chain all molecules donate a single hydrogen bond except one molecule at one of the chain ends. In contrast to all other molecules of the chain which have a single dangling OH bond, this molecule has two. Thus, it has more freedom to move and a higher contribution to the entropy of the chain than the other molecules. If we generate an L-defect,  which donates two hydrogen bonds without accepting any, both molecules at the chain ends have two dangling OH, thus conserving the number of dangling OH bonds. The situation for the D-defect is similar and thus the number of dangling OH bonds is conserved for any number of L- and D-defects in the chain. Only if a hydrogen bond is broken/formed, the number of dangling OH bonds is increased/decreased by one.

Since the entropic contribution accounts for the difference in the phase space volume contributions of a dangling OH bond and one that donates a hydrogen bond, we can obtain an estimate from simple geometric considerations. We assume that an OH bond of a water molecule within a segment that donates a hydrogen bond is restricted to a spherical cap with an opening angle $\alpha$. For a hydrogen bond of a water molecule at the chain end that accepts a single bond, this cap is about half a sphere. The ratio of these areas gives us an estimate for the difference of the contributions to the entropy of these two water molecules.
The surface area of a spherical cap with opening angle $\alpha$ of a sphere with radius $r$ is given by $ A= 2\pi r^2 (1-\cos \alpha) $.
The difference in the entropies between a dangling OH bond and a hydrogen bonded OH bond is given by 
\begin{equation}
\beta \Delta S = \ln \left( \frac{4\pi r^2}{A }\right) = \ln \left( \frac{2}{1-\cos \alpha}\right)
\end{equation}
The entropic contribution is an energy in our effective Hamiltonian and thus equals the negative entropy difference, i.e., $S_\mrm{c}=-\Delta{S}$.
 Table \ref{ENTROPIC} shows results for different opening angles $\alpha$ which are indeed of the order of the entropic contribution, $\beta S_\mrm{c} = -3.96$. 
\begin{table}
\begin{tabular}{c|cccc}
$\alpha$ & $10^\circ$ & $20^\circ$ & $30^\circ$ & $40^\circ$ \\\hline
$\beta \Delta S$ & 4.88 & 3.50 & 2.70 & 2.15  
\end{tabular}
\caption{\label{ENTROPIC}Estimates for the entropic contribution.}
\end{table}


\section{Results}
\label{RESULTS}

The charge picture of the dipole lattice model gives direct physical insight into the microscopic properties of water in nanopores.
The Coulomb-like interactions lead to an attraction between L- and D-defects and also to an attraction between a chain end and the defect next to it. So, the Coulomb interaction has clearly an important influence on these microscopic properties of water in nanopores. The question arises, what is the influence of the Coulomb-like interactions on the overall phase behavior? Or, in other words, which aspects of the system behavior are captured by the approximations that neglect Coulomb-like interactions (NCA and SCA)?

\subsection{Drying/Filling Transition}

\begin{figure}[ht!]
        \epsfig{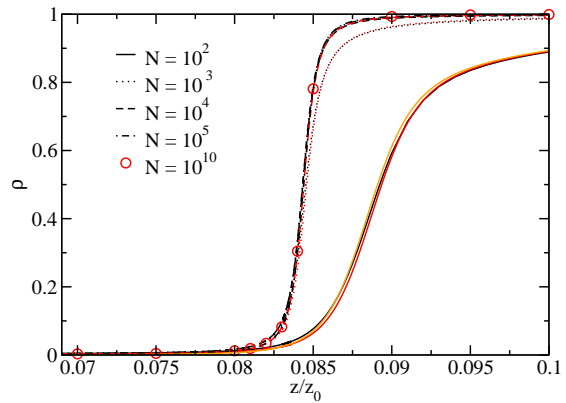}
\caption{\label{MEAN_NSO}Particle density of the full Hamiltonian (black) and the SCA (red). The orange line shows results of NCA for $N=100$. }
\end{figure}

\begin{figure}[ht!]
        \epsfig{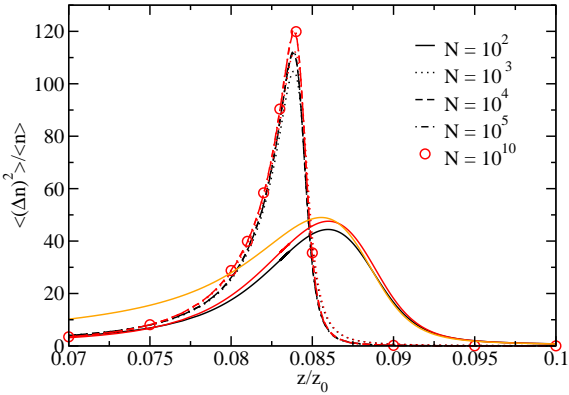}
\caption{\label{RELVAR} Relative variance of the particle number of the full Hamiltonian (black) and the SCA (red). The orange line shows results of NCA for $N=100$.}
\end{figure}

\begin{figure}[ht!]
        \epsfig{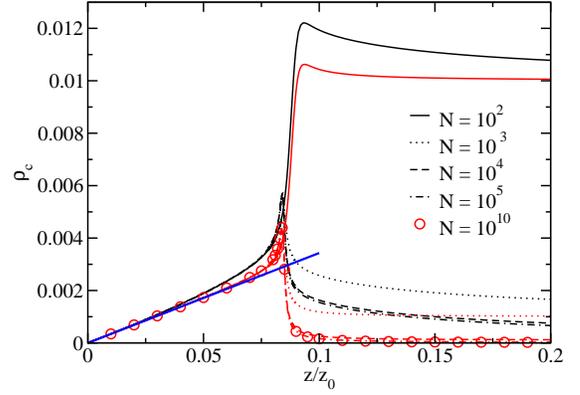}
\caption{\label{CHAINDENS}Chain density of the full Hamiltonian (black) and the SCA (red). The blue line shows the density of non-interacting particles.}
\end{figure}

\begin{figure}[ht!]
        \epsfig{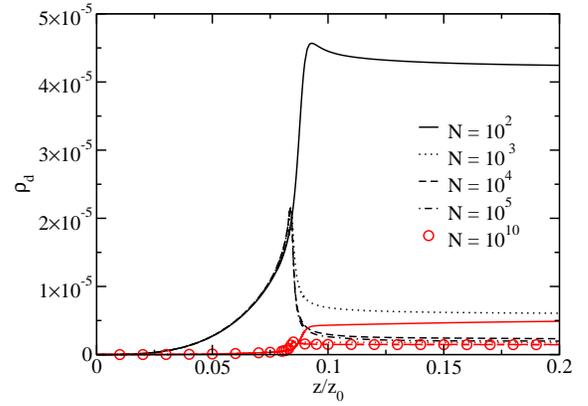}
\caption{\label{DEFDENS} Defect density of the full Hamiltonian (black) and the SCA (red).}
\end{figure}

To clarify the role of the Coulomb-like interactions, we compare results derived in the SCA with results of the full Hamiltonian for the filling transition.  We characterize the system by the particle density, $
\rho=n/N$, (Fig.~\ref{MEAN_NSO}), the particle fluctuations (relative variance, Fig.~\ref{RELVAR}), the chain density  $\rho_\mrm{c}=n_\mrm{c}/N$, i.e., the number of chains per site, (Fig.~\ref{CHAINDENS}), and the defect density $\rho_\mrm{d}=n_\mrm{d}/N$, i.e., the number of defects per site (Fig.~\ref{DEFDENS}), as a function of the relative fugacity.
The results for system sizes $N$=$10^2$, $10^3$, $10^4$, and $10^5$ sites are obtained by reweighting of biased sampling simulation data. Additionally, we show results for the SCA for $N=10^{10}$ from Monte Carlo simulations at the corresponding fugacity values using the effective Hamiltonian of Eq.~(\ref{HEFF}).

The adsorption isotherms (i.e., the average particle density) in Fig.~\ref{MEAN_NSO} and the relative fluctuations of the particle number in Fig.~\ref{RELVAR} show that the results for the SCA are in excellent agreement with the results for the full Hamiltonian for the system sizes studied here.  We find that both the adsorption isotherms and the relative variance are nearly converged to their thermodynamic limits for a system size of $10^4$ sites which is supported by the results for SCA for $N=10^{10}$. 

The adsorption isotherm become steeper with increasing system size but the slope remains finite even in the thermodynamic limit. The relative variance is peaked at the filling transition  as shown in Fig.\ \ref{RELVAR}. Even though the peak height initially grows with increasing system size, it eventually converges to a finite value. This is in agreement with the impossibility of a true first-order phase transition in one dimension for $1/r^3$ interactions \cite{RuelleCommMathPhys1968}.

Figures \ref{MEAN_NSO} and \ref{RELVAR} also show results for the NCA for a system size $N=100$. The adsorption isotherms are in good agreement, but the relative variance decays too slowly for fugacities below the filling transition compared to the results of the full Hamiltonian. The SCA reproduces the results of the full Hamiltonian well, as the system at low fugacities consists mainly of chains of length one which are correctly described in the SCA. 

Since the adsorption isotherms for the NCA and the full Hamiltonian are in good agreement we use this approximation to obtain an estimate for the chemical potential $\mu_{1/2}$, where the system is half full. Assuming that no defects exists, and that the number of particles is much larger than the number of chains we obtain a Hamiltonian that only depends on the particle number, 
\begin{equation}
\mcl{H}'  = n c - \mu n\, ,
\end{equation}
where we treat the chemical potential like a magnetic field in the Ising model and put it in the Hamiltonian.
The $n$ particles do not interact with each other but couple to the field $c-\mu$.  The canonical partition function of this ideal lattice gas in an external field is given by $Z=(1+e^{-\beta(c-\mu)})^N$. The density is obtained by $\rho=\langle n\rangle /N = - 1/Z \partial Z/\partial (\beta \mu)$ which gives $\rho=e^{-\beta(c-\mu)}/(1+e^{-\beta(c-\mu)})$. If we demand that the system is half full, i.e.\ , the density is $\rho=1/2$, then the energy of putting a particle into the system is equaled by the chemical potential, $\mu=c$, which leads to 
\begin{equation}
        \label{MUHALF}
 \mu_{1/2} = 1+E_\mrm{c}-\zeta(3)\, . 
\end{equation}
The corresponding relative fugacity $z_{1/2}/z_0\approx 0.0841$ is in good agreement with the simulation result $z_{1/2}/z_0\approx 0.084$, corresponding to $8.4\%$ relative humidity.

This result can also be derived for a lattice gas with dipole-dipole interactions and exploiting the isomorphism to the Ising model (see App.~\ref{ISO}).

Observables depending on the particle number are well reproduced in the SCA. We obtain insight into the structure of the system by looking at the chain and defect density with changing chemical potential.
We find that at the filling transition both the chain density in Fig.~\ref{CHAINDENS} and the defect density in Fig.~\ref{DEFDENS} are peaked.  The peak in the chain density is also reproduced by the SCA although the curves are below the results of the full Hamiltonian. Thus, the peak in the fragment density reflects the entropic gain through fragmentation for low particle densities.

Figures \ref{CHAINDENS} and \ref{DEFDENS} also show that for the full Hamiltonian the defect density roughly mirrors the fragment density. The reason is, that on the one hand a chain with a defect can lower its free energy by splitting the chain into two at the defect site.
In contrast, the generation of defects costs less energy at the ends of water chains and the number of ends is proportional to the number of fragments. The latter is a consequence of the Coulomb-like attraction between the charge at the chain end and charges forming a defect next to this end. Thus, the results of the SCA do not show a peak in the defect density and do not reproduce the results of the full Hamiltonian at the filling transition even qualitatively.

For fugacities $z$ below the filling transition, the average fragment number decays approximately linearly with $z$ towards zero.  At low fugacity, the system behaves ideally. The grand-canonical partition function of ideal particles is given by $Z=\sum_n{N \choose n} (2 z e^{-\beta S_\mrm{c}})^n$ which leads for small fugacities to a density $\rho \approx 2 z \exp(-\beta S_\mrm{c})$, where $S_\mrm{c}$ is the entropic correction.  The fragment number is then approximately equal to the number of particles, explaining the observed linear decay for small $z$.  Moreover, with fragments too short to carry a defect, the average defect number decays even faster.

\begin{figure}[h]
        \epsfig{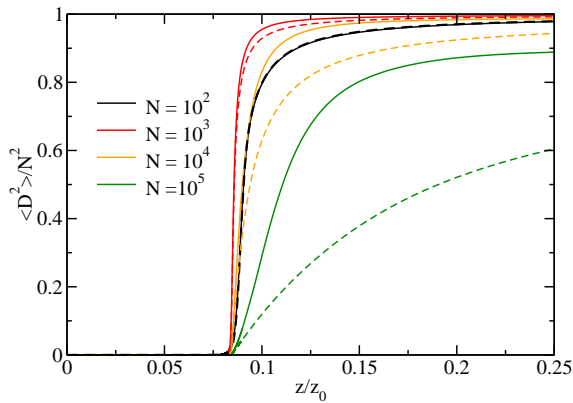}
        \caption{\label{DSQR} The average value of the total dipole moment squared divided by the system size squared as a function of the relative fugacity for different system sizes for the full Hamiltonian (solid lines) and for the SCA (dashed lines).}
\end{figure}

The increasing fragmentation while approaching the filling transition is also the reason for the faster decay of the orientational order for larger system sizes. As measure for the order we use the average of the total dipole moment squared divided by the system size, $\langle D^2\rangle/N^2$ as shown in Fig.~\ref{DSQR} for different system sizes as a function of the relative fugacity. This order parameter is close to unity if the system is \index{}orientationally and translationally ordered and approaches zero for increasing disorder.
Although the filling transition is steepest for $N=10^5$ the order parameter decreases the fastest.
This is direct consequence of the peak in the chain density and the larger gaps for this system size which leads to a weaker coupling of the chains. 

This conclusion is supported by results for the SCA (also shown in Fig.~\ref{DSQR}) which agree nicely with results for the full Hamiltonian for small system sizes of $N=100$ and $N=1000$ where fragmentation plays a minor role. For larger system sizes we observe that the square of the total dipole moment is lower for the SCA then for the full Hamiltonian. This indicates, that for the full Hamiltonian chains are coupled to their next neighbors via Coulomb-like interaction. Since the SCA is lacking these interactions, chains are uncorrelated leading to a lower expectation value for the total dipole moment squared. 

\subsection{Bistability} 
\label{BI}

\begin{figure}
        \includegraphics[width=0.48\textwidth]{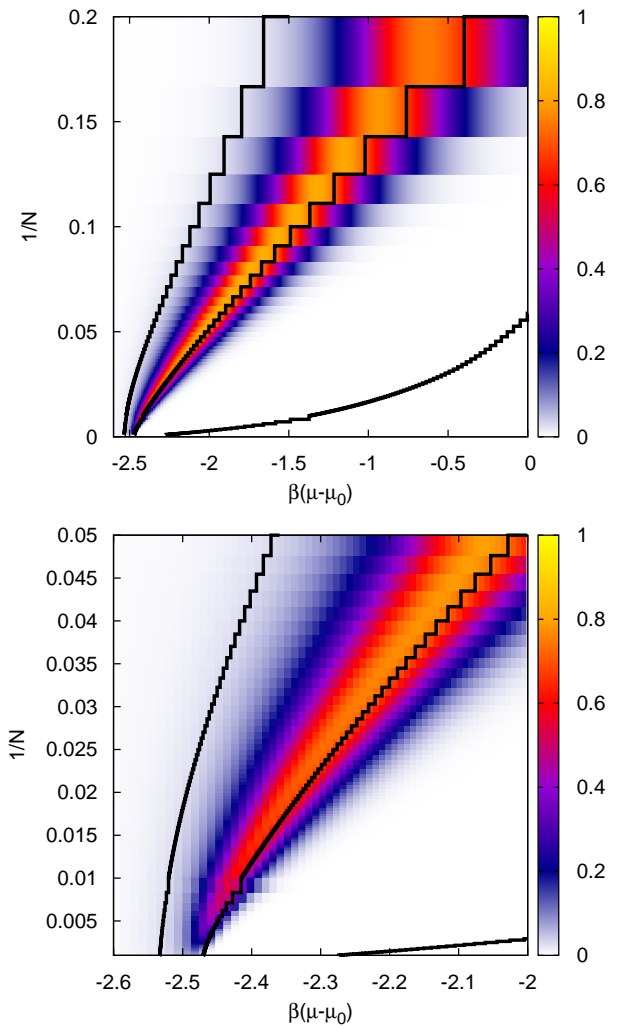}
\caption{\label{bistable}Bistability map. Density plot of the bistability measure $M$ of Eq.~(\ref{BIMEAS}), which is close to one for a bistable system, for all investigated system sizes (top) and an enlarged view for large systems (bottom).  The black solid lines are lines of constant density $\rho=0.01, 0.5$ and $0.99$ from left to right.}
\end{figure}

It was already observed for small systems at ambient conditions that the particle number distribution function is bimodal \cite{HummerNoworytaNature2001,MaibaumChandlerJPCB2003}, showing one peak for the empty and one for the full tube. In the following we investigate how this bistable behavior changes with system size and chemical potential and discuss the influence of the Coulomb-like interactions.

Similar to Maibaum and Chandler in Ref.~\cite{MaibaumChandlerJPCB2003} we start the discussion with the Hamiltonian of a one dimensional lattice gas in an external field $h$,
\begin{equation}
 H = -J \sum_{i=1}^{N-1} s_i s_{i+1} - h \sum_{i=1}^{N} s_i
\end{equation}
where the occupation number of site $i$ is given by $s_i=0,1$ and the coupling constant of occupied sites by $J>0$. This Hamiltonian can be written as 
\begin{equation}
 H=-(J+h) n + J n_\mrm{c} 
\end{equation}
where $n$ is the total occupation number and $n_\mrm{c}$ the number of domains (or chains) of particles. 
The above Hamiltonian is of the form 
\begin{equation}
 H=\mcl{K} n +\mcl{I} n_\mrm{c}
\end{equation}
where $\mcl{K}$ is the coupling constant and $\mcl{I}$ the interface energy. The latter is the energy needed to break a chain into two and form two new interfaces between occupied and empty sites. In case of the NCA of the dipole model, the coupling constant is given by $\mcl{K}= \beta c- \ln z$ and the interface energy by $\mcl{I}=\beta c_\mrm{c}-\ln 2$.

The particle number distribution of an ideal lattice gas (i.e.\ $\mcl{I}=0$) is given by the binomial distribution and therefore shows only a single peak. Only with a positive interface energy bistability of a partly or completely filled system and the empty system is observed. Then the particle number distribution function is given by 
\begin{equation}
 P(n)\propto e^{-\beta \mcl{K} n} \sum_{n_\mrm{c}} \Gamma(n, n_\mrm{c}) e^{-\beta n_\mrm{c} {I}}
\end{equation}
where $\Gamma(n, n_\mrm{c})$ is the number of states depending on the particle number and the number chains, and a functional form that depends on the boundary conditions.

For periodic boundary conditions the number of states is given by 
\begin{equation}
\Gamma(n, n_\mrm{c})={n-1 \choose n_\mrm{c}-1}{N-n \choose n_\mrm{c}}+{N-n-1 \choose n_\mrm{c}-1}{n \choose n_\mrm{c}} \, .
\end{equation}

For a constant number of chains, this function is unimodal and symmetric with respect to the location of its maximum at $n=N/2$. Thus the particle number distribution function is unimodal if one excludes the empty state. Free rather than periodic boundary conditions introduce a small asymmetry in the number of states, i.e.\ 
\begin{equation}
\Gamma(n, n_\mrm{c})={n-1 \choose n_\mrm{c}-1}{N-n+1 \choose n_\mrm{c}} \quad, 
\end{equation}
 which is not sufficient to change this behavior.

\begin{figure}[h]
        \epsfig{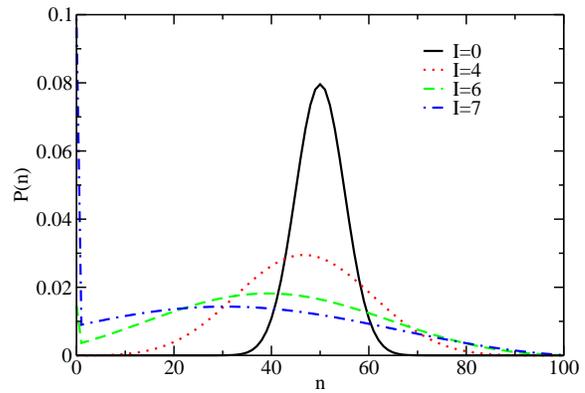}
        \caption{\label{BINOM} Particle number distributions for different values of the interface energy for a system with free boundary conditions of size $N=100$ close to the filling transition ($\mcl{K}=0$). For an interface energy $I\gtrsim6$ we see a second peak for the empty system ($n=0$).}
\end{figure}

For a positive interface energy all non-empty states are energetically penalized according to their number of chains compared to the ideal lattice gas. Thus, the weight of the empty state in the partition function increases relative to the non-empty states and a second peak for $n=0$ appears (see Fig.~\ref{BINOM}). 

In contrast to this, water in nanopores shows a low density peak at densities larger than zero. The reason is that the Coulomb-like interaction of charges of opposite sign at the ends of short ordered chains lowers the internal energy of such chains compared to the energy they would have with only nearest-neighbor interactions. This is already the case for the SCA, where chains of length one, i.e., single dipoles for defect free systems, are treated separately.
In the following, we quantify this bistable behavior for the SCA as a function of the system size and the fugacity.
We calculated the particle number distribution function for system sizes up to $N=1000$ for a certain value of the fugacity $z$.
The particle number distribution function for the SCA is given by
\begin{equation}
 P(n) \propto z^n \mcl{Z}^{(1)}_N(n) 
\end{equation}
with the canonical partition function $\mcl{Z}^{(1)}_N(n)$ given by Eq.~(\ref{CAN_Z_SRA_too}).
We generated particle number distribution functions for fugacities in the range $z/z_0\in[0,1]$ by reweighting.

Bimodal particle number distribution functions show a low-density peak that is given by the binomial distribution of non-interacting particles 
\begin{equation}
 P_\mrm{L}(n) = \frac{1}{(1+z 2 e^{-\beta S_\mrm{c}})^N}{N\choose n} (2 z e^{-\beta S_\mrm{c}})^n
\end{equation}
Thus, we identify the low-density peak as 
\begin{equation}
H_\mrm{L}(n)=P_\mrm{L}(n)\frac{P(0)}{P_\mrm{L}(0)}
\end{equation}
where the factor $P(0)/P_\mrm{L}(0)$ guarantees that the distribution function $P(n)$ and the low-density peak $H_\mrm{L}(n)$ coincide for the empty tube, i.e., $n=0$. The second peak is a high-density, peak defined by $H_\mrm{H}(n)=P(n)-H_\mrm{L}(n)$.

Next, we define a measure $M$ that quantifies the extent of bistability. It should be large if the areas below the two peaks, given by $N_\mrm{H}=\sum_n H_\mrm{H}(n)$ and $N_\mrm{L}=\sum_n H_\mrm{L}(n)=1-N_\mrm{H}$, are of comparable size. Thus, we form the product of the two areas 
\begin{equation}
 \mcl{A} = {4}{N_\mrm{H} N_\mrm{L}}
\end{equation}
The factor $4$ ensures that $\mcl{A}=1$ if the two peaks have equal weight. If one peak dominates the particle number distribution function then $\mcl{A}\approx0$.

If the two peaks were due to coexistence at a first order phase transition, changing the fugacity would only change the relative weight of the two peaks. Here, the location of the peaks changes with the fugacity. Thus, our measure should also include the distance between the two peaks.  
The positions of the peaks are given by the mean values of the low-density and the high-density peak and their distance $R$ can be written as 
\begin{equation}
 R=\frac{1}{N} \sum_n n \left[ \frac{H_\mrm{H}(n)}{N_\mrm{H}}-\frac{H_\mrm{L}(n)}{N_\mrm{L}}\right]
\end{equation}

Thus, we define our measure for bistability $M$ as the product of $\mcl{A}$ and the distance $R$ 
\begin{equation}
\label{BIMEAS}
M = R \mcl{A} = 4 R {N_\mrm{H} N_\mrm{L}} 
\end{equation}

Figure~\ref{bistable} shows a density plot of this measure as a function of the inverse system size and the chemical potential. For small systems, the bimodal structure of the particle number distribution function can be seen for ambient conditions corresponding to $\mu=\mu_0$, as was also observed in computer simulations \cite{HummerNoworytaNature2001}. For larger system sizes the range of the chemical potential where bimodality is observed becomes narrower and the bimodality itself weaker. For $N\gg1000$ bimodality vanishes completely. Also shown are the lines where the system is empty (average density $n/N=\rho=0.01)$, full ($\rho=0.99$), and half filled ($\rho=0.5$). The maximum of the bistability measure is always to the left, i.e., at lower chemical potential, of the line of half filling. The bistability map for NCA (not shown) is nearly identical to the map for SCA.

\begin{figure}
        \includegraphics[width=0.45\textwidth]{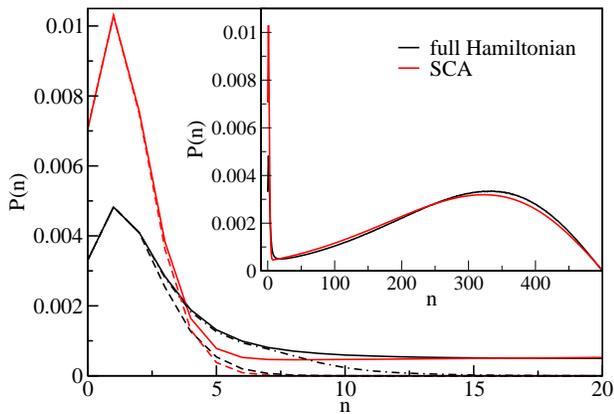}
\caption{\label{DIST}Particle distribution functions for $N=500$ and $z/z_0=0.085$ for the full Hamiltonian (solid, black) and in the SCA (solid, red). The dashed lines show fits of the low density peaks with binomial distributions of non-interacting single dipoles. The dashed-dotted line is a fit of a binomial distribution of non-interacting chains (with charges at their ends) up to a length of ten particles. The inset shows the distribution functions for the full range of the particle number. }
\end{figure}

The situation for the full Hamiltonian is slightly different to that for the SCA as exemplified in Fig.~\ref{DIST}.
The low-density peak is not perfectly reproduced by a binomial distribution of non-interacting, single dipoles. Instead, we have to take into account the contributions of longer chains, that do not interact with each other, but whose energy is lowered (compared to the SCA) due to the effective charges at their ends. The peak is well reproduced if we include chains with lengths up to ten sites.
Thus, the form of the low density peak can be explained by the lower energies of chains in the full Hamiltonian compared to the SCA.

\section{Summary and Conclusions}

\label{DISCUSS}

We have studied a one-dimensional dipole model in various equivalent representations, and also in approximate formulations.
The dipole model is a simple and powerful description of water in nanopores.  It allows us to characterize the phase-like properties of confined water up to macroscopic dimensions \cite{KoefingerDellagoPNAS2008}. Here we have explored the filling thermodynamics by calculating adsorption isotherms and particle number fluctuations as a function of relative humidity.  We have also characterized the regime of bistability, with distinct gas-like and liquid-like states, as a function of tube length and fugacity. 

The model can be represented in terms of individual dipoles, ordered segments, or effective charges with Coulomb-like interactions.  The segment picture is not only an essential step in the derivation of the charge representation from the dipole picture; it also results in a simple characterization of the configuration space and thus an efficient formulation of non-local trial moves for Monte Carlo simulations. The charge picture is the physically most appealing representation as the long-range interactions are due to charges at the ends of chain segments. It also reduces the computational cost of calculating the Hamiltonian.

In the charge picture, the Coulomb-like effective interactions of defects and chain ends result from effective charges, whose magnitude depend on the dipole moment of water molecules and their distance in the water wire. L- and D- defects carry effective charges of opposite sign with a magnitude that is twice that of the charges at the chain endpoints. An excess proton in the chain, corresponding to an L-defect with an extra proton, carries an effective charge that is reduced considerably with respect to the charge of the bare proton. This polarization effect is important for the free energetics of proton transfer through water-filled narrow pores \cite{DellagoHummerPRL2006}. 

We introduced two approximate representations of the dipole model.  By neglecting the long-range interaction in the charge representation, one arrives at a Hamiltonian that only depends on the particle number, the chain number, and the defect number. This approximation can be improved by treating charge interactions within segments of length one explicitly. The resulting approximation produces accurate descriptions for observables depending on the numbers of particles and fragments. For this so-called singlet-charge-approximation we can count the number of states as a function of the independent variables. This model can thus be formulated in terms of an effective Hamiltonian with a phase-space of only four dimensions. 

We used these approximations to study the bistability of the particle number distribution function, which gets weaker with increasing system size. The biggest effect of the long-range Coulomb-like interactions, as compared to the short-range interactions in an Ising-like lattice gas, is to lower the energies of short chains.  This energy lowering accounts for the low-density peaks in the particle-number distributions, with small but non-zero particle numbers. Nevertheless, for system sizes that are not too small, the bistable behavior of water in nanopores is captured by the SCA model without long-range interactions.

The dipole model introduced here is general and should be applicable to other quasi-one dimensional systems with dipolar interactions, including polar fluids other than water as well as magnetic nanoparticles and colloids \cite{ShevchenkoMurrayNature2006,PuntesAlivisatosApplPhysLett2001}. It should also prove useful in studies of three-dimensionally packed arrays of one-dimensional chains, such as those formed in membranes of parallel nanochannels \cite{KalraHummerPNAS2003,KoefingerDellagoPNAS2008}.

\begin{acknowledgments}
J.K.\ and C.D.\ acknowledge support from the Austrian Science Fund (FWF) under Grant No.\ P17178-N02 and within the Science College "Computational Materials Science" under grant W004, and from the University of Vienna through the 
University Focus Research Area {\em Materials Science} (project ``Multi-scale Simulations of Materials Properties and Processes in Materials'').
J.K.\ thanks the European Science Foundation (ESF) for a short visit grant within the activity SimBioMa: `Molecular Simulations in Biosystems and Material Science' and the NIH for additional support. G.H. was supported by the Intramural Research Program of the NIH, NIDDK.
\end{acknowledgments}

\begin{appendix}
\section{Number of States}
\label{app:NRSTATES}
For the NCA and the SCA we can count the number of states as a function of the independent variables of the respective Hamiltonians. Here, we show the derivation for the SCA, where the number of states depends on the particle number $n$, chain number $n_\mrm{c}$, defect number $n_\mrm{d}$, and number $n_\mrm{I}$ of single dipoles corresponding to segments of length one. For the simpler NCA, the derivation of the number of states is similar.

To model the free boundary conditions we add to our fully occupied system consisting of $N$ sites an empty section of length one at each end. Thus, the fully occupied system is given by an empty section of length one, an occupied section of length $N$, and again an empty section of length one. These end sections are useful for the formulation of trial moves and for the derivation of the number of states, as we will see in the following.

To calculate the number of states we first take a section of length $n-n_\mrm{d}-n_\mrm{I}$ and split it into $n_\mrm{s}-n_\mrm{I}$ parts, each of which is at least of length two, i.e., $l_\mrm{min}=2$. 
To split a section of $M$ sites into $m$ parts we have to choose $m-1$ of the $M-1$ points between sites where the section can be split. The number of possibilities to do so is given by the binomial coefficient ${M-1\choose m-1}$. If we demand that each new section consists of at least $l_\mrm{min}$ sites then the number of points where we can split the section into parts is reduced by $(l_\mrm{min}-1) m$. 
Thus, the number of possibilities, $\gamma$, of splitting a section with $M$ sites in $m$ sub-sections, where each section has a length $l\geq l_\mrm{min}$ is given by
\begin{equation}
\label{APPENDIXGAMMA}
\gamma={M-(l_\mrm{min} -1)m-1 \choose  m-1}\, .
\end{equation}
Inserting $M=n-n_\mrm{d}-n_\mrm{I}$, $m=n_\mrm{s}-n_\mrm{I}$, and $l_\mrm{min}=2$ in the above equation we obtain the first binomial coefficient of Eq.~(\ref{GAMMA1}).

Next we count in how many ways we can combine the $n_\mrm{s}-n_\mrm{I}$ segments with lengths larger than one and the $n_\mrm{I}$ segments of length one to a particular sequence of these $n_s$ segments. We do so by choosing positions for the $n_\mrm{I}$ identical, single dipoles out of $n_\mrm{s}$ possible positions,  which gives the second binomial coefficient of Eq.~(\ref{GAMMA1}). Then, we put the remaining $n_\mrm{s}-n_\mrm{I}$ segments of lengths larger than one on the remaining positions without changing their order, which is uniquely determined.

Grouping these segments to chains corresponds to splitting this sequence of $n_\mrm{s}$ segments into $n_\mrm{c}$ parts, each consisting of at least one segment. This gives the third binomial coefficient of Eq.~(\ref{GAMMA1}). Each of these $n_c$ chains has two possible orientations which gives the factor $2^{n_\mrm{c}}$.

Finally, $N-n+2$ empty sites have to be partitioned in $n_\mrm{c}+1$ sections that are at least of length one, i.e., $l_\mrm{min}=1$ in Eq.~(\ref{APPENDIXGAMMA}), which gives the last Binomial coefficient of Eq.~(\ref{GAMMA1}). By alternating empty sections and chains a particular configuration is obtained, which is again uniquely determined.

To do Monte Carlo simulation of the SCA effective Hamiltonian given by Eq.~(\ref{HEFF}), we need not only the degeneracy, but we also have to know the limits of the volume spanned by the variables $\{n, n_\mrm{d}, n_\mrm{c}, n_\mrm{I}\}$, i.e., their minimum and maximum values. The minimum particle number is $n^\mrm{min}=0$ and the maximum particle number is $n^\mrm{max}(N)=N$.
The minimum number of defects is $n_\mrm{d}^\mrm{min}=0$ and the maximum number is given by $n_\mrm{d}=0$ for $n<3$ and otherwise
\begin{equation}
n_\mrm{d}^\mrm{max}(n)= \left\{\begin{array}{ccl}\frac{n}{2}-1 &\mbox{for}& n\ \mbox{even}\\ \frac{n-1}{2}&\mbox{for}& n\  \mbox{odd}\end{array}\right. .
\end{equation}
The minimum number of chains is 
\begin{equation}
 n_\mrm{c}^\mrm{min}(n) = \left\{\begin{array}{ccl}0 & \mbox{for} & n=0 \\ 1 &\mbox{for} & n>0\end{array}\right.
\end{equation}
and the maximum number is  $n_\mrm{c}^\mrm{max}(N,0,n_\mrm{d})=0$ for the empty system and otherwise
\begin{eqnarray}
 n_\mrm{c}^\mrm{max}(N,n,n_\mrm{d}) =\min\left\{N-n+1, n-2n_\mrm{d}\right\} .
\end{eqnarray}
The minimum number of chains of length one is given by 
\begin{equation}
 n_\mrm{I}^\mrm{min} (n, n_\mrm{d}, n_\mrm{c})= \left\{\begin{array}{ccl}0 & \mbox{for} & n-n_\mrm{d}\geq2n_\mrm{s} \\ 2 n_\mrm{s}-n+n_\mrm{d} &\mbox{for} & n-n_\mrm{d}<2n_\mrm{s}\, .\end{array}\right.
\end{equation}
The maximum number of chains of length 1 is given by 
\begin{equation}
 n_\mrm{I}^\mrm{max} (n, n_\mrm{d}, n_\mrm{c})= \left\{\begin{array}{ccl}n_\mrm{s} & \mbox{for} & n-n_\mrm{d}=n_\mrm{s} \\ n_\mrm{s}-1 &\mbox{for} & n_\mrm{s}<n\, .\end{array}\right. 
\end{equation}

\section{Trial Moves}
\label{TRIAL}

In the following we present the trial moves for the Monte Carlo simulation of the dipole lattice model.

In the Metropolis algorithm \cite {FrenkelSmit2002} the transition probability from an old state $o$ to a new state $n$ is  given by the product of the generation probability of a move, $P_\mrm{gen}(o\rightarrow n)$ and the acceptance probability, $P_\mrm{acc}(o\rightarrow n)$.
Imposing detailed balance, the Metropolis acceptance probability in the canonical ensemble is given by 
\begin{equation}
\label{ACCPROP}
 P_\mrm{acc}(o\rightarrow n) =\min\left\{1,\frac{P_\mrm{gen}(n \rightarrow o)e^{-\beta  E(n)}}{P_\mrm{gen}(o\rightarrow n)e^{-\beta E(o)}}\right\}
\end{equation}
and correspondingly for $P_\mrm{acc}(n\rightarrow o)$. 
In simulations of the grand-canonical ensemble, the energies contain additional terms $-n \ln z$ where $n$ is the fluctuating particle number.
Usually the generation probabilities of the forward and the backward move are chosen to be equal and they cancel each other in the above equation. 

In our simulation, a configuration is given in the segment picture, i.e., by the lengths of all sections and their orientations. This is a simple way to include the configurational constraints mentioned above but has the disadvantage that for some trial moves the generation probability for the forward and the backward move is asymmetric. These asymmetric generation probabilities have to be explicitly included in Eq.~(\ref{ACCPROP}). This is the case for defect generation and recombination, chain splitting and joining, and the insertion and removal of a single dipole, as is explained below.

For simplicity, we use in the remaining part of this section the word ``choose'' when we mean ``choose with equal probability'', i.e., when we draw some quantity from a uniform distribution.

For the {\it displacement of a defect} we choose a chain $c\in \{1, \ldots, n_\mrm{c}\}$ that consists of $m$ segments from which we choose segment $i \in \left\{1, \ldots, m-1\right\}$. We change the length of segment $i$ by $d\Delta l$ and the length of segment $i+1$ by $-d\Delta l$ where we have chosen a length $\Delta l \in \{1, \Delta_\mrm{max}\}$ and a direction $d=\pm1$. The generation probability is given by
\begin{equation}
 P^\mrm{dis}_\mrm{gen}=\frac{1}{2 n_\mrm{c}(m-1)\Delta_\mrm{max}}\quad .
\end{equation}

For the {\it generation of a defect} we choose a chain $c\in \{1, \ldots, n_\mrm{c}\}$ that consists of $m$ segments from which we choose segment $i \in \left\{1, \ldots, m \right\}$. If the length of this segment, $l$, is long enough to carry a defect, i.e., $l\geq3$, then we choose a length $l' \in \{1,\ldots, l-2\}$ and a direction $d=\pm1$. The segment $i$ is split in two chains of length $l_1=l'$ and $l_2=l-l'$ and all chains of fragment $f$, on the side given by $d$ are reoriented. 
The generation probability is given by 
\begin{equation}
 P^\mrm{gen}_\mrm{gen}=\frac{1}{2 n_\mrm{c}m (l-2)}\quad .
\end{equation}
This move increases the number of defects $n_\mrm{d}$ by one.
 
For {\it defect recombination} we choose a chain $c\in \{1, \ldots, n_\mrm{c}\}$ that consists of $m$ segments from which we choose segment $i \in \left\{1, \ldots, m-1\right\}$. Additionally we choose a direction $d=\pm1$ and join segments $i$ and $i+1$ to a new segment with length $l=l_i+l_{i+1}+1$. All segments on the side $d$ are reoriented. The generation probability is given by 
\begin{equation}
 P^\mrm{rec}_\mrm{gen}=\frac{1}{2 n_\mrm{c}(m-1)}\quad .
\end{equation}
This moves decreases the number of defects $n_\mrm{d}$ by one.

For the {\it displacement of a fragment} we choose a chain $c\in \{1, \ldots, n_\mrm{c}\}$, a direction $d=\pm1$, and a displacement $\Delta l \in \{1, \ldots, \Delta_\mrm{max}\}$. The empty section on the side $d$ of the fragment is lengthened by $\Delta l$ and the empty section on the opposite side is shortened by $\Delta l$.
The generation probability is given by 
\begin{equation}
 P^\mrm{fra}_\mrm{gen}=\frac{1}{2 n_\mrm{c} \Delta_\mrm{max}}\quad .
\end{equation}

The generation probability for the {\it reorientation of a chain}, i.e., the reorientation of all segments of chain $c\in \{1, \ldots, n_\mrm{c}\}$, is given by
\begin{equation}
 P^\mrm{reo}_\mrm{gen}=\frac{1}{n_\mrm{c}}\quad .
\end{equation}

The {\it exchange} move shortens a segment $i\in \{1, \ldots, m_1\}$ of fragment $c_1$ consisting of $m_1$ segments, and lengthens a segment $j\in \{1, \ldots, m_2\}$ of $c_2$ consisting of $m_2$ segments a length $\Delta l\in \{1, \Delta_\mrm{max}\}$, therefore conserving the number of occupied sites. The generation probability is given by 
\begin{equation}
 P^\mrm{exc}_\mrm{gen}=\frac{1}{n_\mrm{c}^2 m_1 m_2}\quad .
\end{equation}

To {\it split a chain in two chains} we choose a chain $c_1\in \{1, \ldots, n_\mrm{c}-1\}$ and a segment $i \in \left\{1, \ldots, m\right\}$ of length $l$. Next we have to choose where in the segment $i$ a bond is broken by choosing a length $l'\in \{1, l-1\}$. One of the new fragments is displaced ($d=\pm1$) a length $\Delta l \in \{1, \Delta_\mrm{max}\}$. The generation probability is given by
\begin{equation}
 P^\mrm{spl}_\mrm{gen}=\frac{1}{2 n_\mrm{c} m (l-1) \Delta_\mrm{max}}
\end{equation}
This move increases the number of chains by one and the number of segments by one.

The inverse move is the {\it joining of two chains} to  a single chain. We choose a chain $c\in \{1, \ldots, n_\mrm{c}-1\}$. If the last ordered segment of chain $c$  and the first ordered segment of chain $c+1$ have the same direction and if they are not further apart than $\Delta_\mrm{max}$ (i.e., the length of the empty section between them is $l\leq\Delta_\mrm{max}$) then we try to join them. We choose a direction $d=\pm1$ that decides if the left chain is moved towards the right or the right towards the left. We get for the generation probability
\begin{equation}
 P^\mrm{joi}_\mrm{gen}=\frac{1}{2 (n_\mrm{c}-1)}\quad .
\end{equation}
This move decreases the number of chains by one and the number of ordered segments by one.

Next we present moves that change the occupation number.
The {\it transfer move} adds or removes dipoles at the end of chains. First we choose a chain $c\in\{1, \ldots, n_\mrm{c}\}$, at which end particles are transfered ($d_1=\pm1$), and how many particles ($\Delta l\in \{1, \Delta_\mrm{max}\}$) are either added or removed by lengthening or shortening of the chosen end segment ($d_2=\pm1$). Lengthening of the end segment is only possible if the empty section next to it is longer than the number of added particles. This also guarantees that the number of chains is not changed.
This gives a generation probability of 
 \begin{equation}
 P^\mrm{tra}_\mrm{gen}=\frac{1}{4 n_\mrm{c} \Delta_\mrm{max}}\quad .
\end{equation}

The above move is only applicable if there are already occupied sites.
Therefore, we also {\it insert single dipoles} in empty sections.
To do so we choose an empty section $i \in \{1, \ldots, n_\mrm{c}+1\}$ and a site by choosing a length $ l' \in \{ 1, l-2\}$ for the empty section on the left of the inserted dipole, which we assign an orientation $d=\pm1$. This results in a generation probability
\begin{equation}
 P^\mrm{ins}_\mrm{gen}=\frac{1}{2 (n_\mrm{c}+1) (l-2)}\quad .
\end{equation}
This move increases the occupation number, the number of chains, and the number of ordered segments by one.

The inverse move {\it removes a single dipole} by choosing a chain $c \in \{1, \ldots, n_\mrm{c}\}$ and checking if it is of length one. If so, we remove the single dipole by eliminating this chain consisting of segment $j$ and the empty section to its right with index $j+1$. The empty section to its left gets the new length $l_{j-1}'=l_{j-1}+l_{j+1}+1$.
The generation probability is 
\begin{equation}
 P^\mrm{rem}_\mrm{gen}=\frac{1}{n_\mrm{c}}\quad .
\end{equation}
This move decreases the occupation number, the number of chains, and the number of ordered segments by one.

\section{Biased sampling}
\label{BIAS}
The particle number distribution function $P(n, z')$ at the fugacity $z'$ is obtained from the particle number distribution function $P_w(n)$, that stems from a biased sampling simulation at the fugacity $z$, by unfolding of the weight function $w(n)$ and reweighting to the new fugacity $z'$,
\begin{equation}
 P(n,z')=\frac{1}{N(z')} P_w(n) z^{-w(n)} \left(\frac{z'}{z}\right)^{n}
\end{equation}
with the normalization constant $N(z')$ given by 
\begin{equation}
 N(z') =\sum_{n=0}^{N} P_w(n) z^{-w(n)} \left(\frac{z'}{z}\right)^{n} \quad .
\end{equation}

If we want to calculate the average value of the observable $\mcl O$ (which can be any element of the sampled list or a function of these elements)
 from a biased simulation, we need the joint distribution function of the order parameter and the observable in the biased ensemble given by 
\begin{equation}
P_w(n, \mathcal{O})=\frac{1}{M} \sum_i \delta(n^{(i)}-n)  \delta(\mathcal{O}^{(i)}-\mathcal{O}) 
\end{equation}
for a discrete observable $\mcl{O}$, where $M$ is the number of samples and $\delta(x)$ is Dirac's delta function.
 We obtain the average of the observable $\mcl{O}$ at a fugacity $z'$ by evaluating 
\begin{equation}
 \langle \mcl{O} \rangle = \frac{1}{N(z')}\sum_n \sum_\mcl{O} P_w(n, \mathcal{O}) \mcl{O} z^{-w(n)} \left(\frac{z'}{z}\right)^{n} \, .
\end{equation}

Instead of calculating the two-dimensional histogram and performing the above average, we calculate the following average of the observable $\mcl{O}$ for each value of the order parameter in the biased ensemble, i.e.,
\begin{equation}
 \langle{\mathcal{O}}\rangle_w(n) =  \sum_\mcl{O} P_w(n, \mathcal{O})\mcl{O}= \frac{1}{M} \sum_i \delta(n^{(i)}-n) \mathcal{O}^{(i)} \, .
\end{equation}

We then do the unfolding of the weight function and the reweighting to the new fugacity $z'$ in a single step and obtain for the average of the observable $\mcl{O}$ as a function of the order parameter 
\begin{equation}
 \langle \mcl{O} \rangle(n) =\frac{1}{N(z')} \langle{\mathcal{O}}\rangle_w(n) z^{-w(n)} \left(\frac{z'}{z}\right)^{n} \quad .
\end{equation}
The average value of the observable $\mcl{O}$ at the new fugacity $z'$ is then obtained as $\langle \mcl{O} \rangle =\sum_n\langle \mcl{O} \rangle(n) $.

\section{Lattice Gas}
\label{ISO}
We determine the chemical potential, $\mu_{1/2}$, where the system is half filled by exploiting the isomorphism between the Ising model in an external field in the canonical ensemble and a lattice gas in the grand canonical ensemble. This allows us to determine $\mu_{1/2}$ from symmetry considerations.
The Hamiltonian of the lattice gas is given by 
\begin{equation}
 H = \frac{1}{2} \sum_{i,j} n_i n_j J(|j-i|)
\end{equation}
where $n_i=0,1$ is the occupation number of site $i$ and the interaction potential given by
\begin{equation}
  J(|j-i|) = \left\{\begin{array}{cl}
              0 & \mbox{for}\quad i=j \\
                E_\mrm{c} &\mbox{for}\quad |j-i|=1\\
                - |j-i|^{-3} & \mbox{else} \, .
             \end{array}\right.
\end{equation}
The grand canonical partition function of this lattice gas is given by 
\begin{equation}
\Xi = \sum_{n_k=0,1} \exp\left[-\beta \left(H - \mu \sum_{i} n_i\right) \right] \quad,
\end{equation}
which is isomorphic to the canonical partition function of the Ising model with dipole-dipole interactions
\begin{equation}
Q = \sum_{s_n=-1,1} \exp\left[-\beta \left( \frac{1}{2} \sum_{i,j} s_i s_j \tilde J(|j-i|)- h \sum_{i} s_i\right) \right] 
\end{equation}
with $s_i=1$ corresponding to $n_i=1$, $s_i=-1$ corresponding to $n_i=0$, 
 $\tilde J(|j-i|)=J(|j-i|)/4$, and  $\mu = 4 \tilde J+2h$ with $\tilde J=\sum_{k=1}^{\infty} \tilde J(k)$.
For vanishing external field $h$ the magnetization of the Ising model vanishes which corresponds to a half filled state for the lattice gas.
Thus, $\mu_{1/2}= 4 \tilde J$  which gives 
\begin{equation}
\mu_{1/2} = E_\mrm{c}+[1-\zeta(3)]
\end{equation}
in agreement with Eq.~(\ref{MUHALF}).
 
\end{appendix}

 \end{document}